\documentclass[preprint,12pt]{elsarticle}
\usepackage{amssymb}
\usepackage{amsmath}
\usepackage{amsfonts}
\usepackage{romannum}
\usepackage{titlesec}
\titleformat{\section}{\normalfont\bfseries\scshape}{\thesection}{1em}{}
\usepackage{fancyhdr}

\begin{document}

\begin{frontmatter}

\title{Fermion-boson symmetry and quantum field theory}

\author[a]{Sau Lan Wu}
\author[b]{Tai Tsun Wu\corref{correspondingauthor}}
\cortext[correspondingauthor]{Corresponding author}
\author[a]{Chen Zhou}

\address[a]{Physics Department, University of Wisconsin-Madison,\\ Madison, WI~53706, U.~S.~A.}
\address[b]{Gordon McKay Laboratory, Harvard University, \\Cambridge, MA~02138, U.~S.~A.}

\pagenumbering{arabic}

\begin{abstract}

The application of fermion-boson symmetry to the standard model leads to the following: first, there are three generations of scalar quarks and scalar leptons in addition to the known quarks and leptons, and, secondly, the divergences in the perturbation series for the standard model are reduced. In the light of experimental data from LEP, Tevatron Collider, and LHC, some consequences of these two statements taken together are discussed.

A series of experiments are proposed to search for the scalar quarks and scalar leptons at the Large Hadron Collider. The first step in this search is to look for new fermions by analyzing events with a pair of oppositely changed leptons both with large transverse momenta. The scalar quarks and the scalar leptons are then searched for through their decays into these new fermions plus a known quark or lepton.

\end{abstract}

\end{frontmatter}


\section{Introduction}
\thispagestyle{fancy}
\fancyhf{}
\rhead{CERN-TH-2017-046 \\}

The concept of fermion-boson symmetry has a long history~\cite{goifand}. It's the purpose of the present paper to study this concept from an alternative point of view, where emphasis is placed on high-energy experiments.

Whatever the precise meaning may be for this fermion-boson symmetry, it must imply that, in addition to the well-known quarks and leptons, there are scalar quarks and scalar leptons. Scalar quarks have the same quantum number as the quarks except the spin, and scalar leptons have the same quantum number as the leptons again except the spin.

The behavior of quarks and leptons are well described by the standard model of Glashow, Weinberg, and Salam~\cite{sglashow}. The standard model is a Yang-Mills non-Abelain gauge theory~\cite{CNYRL} with the gauge group $U(1) \times SU(2) \times SU(3)$. The standard model, with the addition of scalar quarks and scalar leptons, is described in Sec.2. From the experimental data from LEP of CERN, some information is available on the masses of these new scalar particles.

When these scalar particles are added to the standard model, there are then, in the perturbative treatment, not only fermion loops due to the quarks and leptons, but also boson loops due to the scalar quarks and scalar leptons. Since the contributions from these two types of loops have opposite signs due to the anti-commutation relations of the fermions, the possibility is entertained that the cancellation of these contributions can lead to the reduction of divergences in the standard model. This is discussed in Sec.3.

When fermion-boson symmetry is combined with this cancellation of divergences, the expected mass ranges of the scalar quarks and scalar leptons are much reduced. This and other consequences of this combination are discussed in the rest of the present paper, including especially a possible new experimental method of searching for the new particles.

\section{Scalar quarks and scalar leptons}

In the standard model, there are three generations of quarks and leptons. They are usually denoted by 

\begin{equation}
\begin{aligned}
u_L, u_R, d_L, d_R; c_L, c_R, s_L, s_R; t_L, t_R, b_L, b_R
\label{aba:eq2.1}
\end{aligned}
\end{equation}

\noindent and 

\begin{equation}
\begin{split}
\begin{aligned}
\nu_{eL}, \nu_{eR}, e_L, e_R; \nu_{\mu L}, \nu_{\mu R}, \mu_L, \mu_R; \nu_{\tau L}, \nu_{\tau R}, \tau_L, \tau_R.
\label{aba:eq2.2}
\end{aligned}
\end{split}
\end{equation}

Using the notation of a tilde to denote a scalar quark or a scalar lepton, the corresponding three generations of scalar quarks and scalar leptons are 

\begin{equation}
\begin{split}
\begin{aligned}
\widetilde{u}_L, \widetilde{u}_R, \widetilde{d}_L, \widetilde{d}_R; \widetilde{c}_L, \widetilde{c}_R, \widetilde{s}_L, \widetilde{s}_R; \widetilde{t}_L, \widetilde{t}_R, \widetilde{b}_L, \widetilde{b}_R.
\label{aba:eq2.3}
\end{aligned}
\end{split}
\end{equation}

\noindent and 

\begin{equation}
\begin{split}
\begin{aligned}
\widetilde{\nu}_{eL}, \widetilde{\nu}_{eR}, \widetilde{e}_L, \widetilde{e}_R; \widetilde{\nu}_{\mu L}, \widetilde{\nu}_{\mu R}, \widetilde{\mu}_L, \widetilde{\mu}_R; \widetilde{\nu}_{\tau L}, \widetilde{\nu}_{\tau R}, \widetilde{\tau}_L, \widetilde{\tau}_R.
\label{aba:eq2.4}
\end{aligned}
\end{split}
\end{equation}

As is well known, the meaning of the subscripts $L$ and $R$ is quite different for fermions and for bosons. For example, while $d_L$ and $d_R$ are the left-handed and right-handed components of the fermion $d$ quark, $\widetilde{d}_L$ and $\widetilde{d}_R$ are two independent scalar quarks. Nevertheless, $\widetilde{d}_L$ and $\widetilde{d}_R$ will be referred to as left-handed and right-handed respectively.

Since $\widetilde{d}_L$ and $\widetilde{d}_R$ are two independent scalar quarks, there is no relation between their masses, i.e., the masses of the scalar quarks $\widetilde{d}_L$ and $\widetilde{d}_R$ are in general different. Therefore, for each generation, there are two different masses for the quarks and two different masses for the leptons, while there are four different masses for the scalar quarks and four different masses for the scalar leptons.

Similar to the cases of quarks and leptons, the left-handed scalar quarks and scalar leptons are $SU(2)$ doublets while the right-handed scalar quarks and scalar leptons are $SU(2)$ singlets.

The experimental data from the $LEP$ $e^+e^-$ collider can be used to give constraints to the masses of the scalar quarks and scalar leptons: the width of the $Z$ is given quite accurately by its coupling to the known quarks and leptons~\cite{zwidth}. This fact implies that every scalar quark and every scalar lepton that couples to the $Z$ must have a mass larger that $\frac{1}{2}m_z$. [Because of experimental error, masses slightly smaller than $\frac{1}{2}m_z$ cannot be excluded. This is especially so because of the $spin-0$ nature of the scalar quarks and scalar leptons.]

Which scalar quark and/or which scalar lepton is not coupled to the $Z$? Since the $Z$ in the standard model is a superposition of the $U(1)$ gauge particle and the neutral component of the $SU(2)$ gauge particle, a scalar quark or a scalar lepton that does not couple to the $Z$ must not have any $U(1)$ quantum number nor any $SU(2)$ quantum number. This means that the only scalar quarks and scalar leptons that do not couple to the $Z$ are the three right-handed scalar neutrinos $\widetilde{\nu}_{eR}$, $\widetilde{\nu}_{\mu R}$, and $\widetilde{\nu}_{\tau R}$.

\section{Divergences in quantum field theory}\label{sec3}

Fermion-boson symmetry is very much related to the divergences in quantum field theory. It is the purpose of this Sec.3 to discuss this relation in some detail.

All the known spin 1 elementary particles are gauge particles, either an Abelian gauge particle or a Yang-Mills non-Abelian gauge particle~\cite{CNYRL}. While the photon for the Abelian case has been known for a century~\cite{einstein}, the first Yang-Mills non-Abelian gauge particle was first observed less than forty years ago at DESY~\cite{slwu}. There are many ways to understand the spin 1 particles being gauge particles. From the point of view of perturbative quantum field theory, here is an especially relevant one: spin 1 particles tend to lead to divergences, and it is gauge invariance that reduces such divergence. 

Such reduction of divergences can also be due to fermion-boson symmetry. Because of the anti-commutation relations for fermion field, a fermion loop has an extra minus sign compared with a boson loop. Because of this minus sign, the presence of fermion-boson symmetry tends to lead to cancellations between the contributions from a fermion loop and a boson loop.

These two ways of reducing divergences in the perturbation calculation of quantum field theory\textemdash gauge invariance and fermion-boson symmetry\textemdash are of fundamentally different nature. While gauge invariance disallows the possibly most divergent term, the presence of both types of loops when there is fermion-boson symmetry implies merely that there is an extra minus sign. This point requires a more detailed discussion.

In order for the cancellation to occur between fermion loops and boson loops to occur, it is not sufficient to have opposite signs; it is also necessary for the magnitudes of the contributions to be equal.

To the best knowledge of the authors, the first paper on such cancellations is that of Ferrara, Girardello, and Palumbo, and this paper was followed shortly thereafter by one of Veltman, and then by a number of later papers~\cite{ferrara}. The results of all these papers take the following form: within the one-loop approximation, the sum of the squares of the fermion masses is equal to the sum of the squares of the boson masses with suitable numerical coefficients.

Independent of the details of the considerations presented in~\cite{ferrara}, the desired cancellation between the fermion loops and the boson loops can be expected to happen only if the fermion masses and the boson masses are comparable. Thus, from the point of view of the quantum field theory for the standard model with the addition of the scalar quarks and scalar leptons as described in Sec. 2, the masses of these scalar quarks and scalar leptons can be expected to be close to these of the the known quarks and leptons.

What does this mean? Among all the known elementary particles, the top quark is the heaviest one. Moreover, since the top quark was discovered experimentally at Fermilab in 1995~\cite{FERB}, it has remained the heaviest for more than twenty years. Its mass of $173$ $GeV/c^2$ is less than twice that of $Z$. Therefore the conclusion is

\begin{equation}
\begin{split}
\begin{aligned}
m_{\widetilde{q}}=O(m_z)
\label{aba:eq3.1}
\end{aligned}
\end{split}
\end{equation}

\noindent and 

\begin{equation}
\begin{split}
\begin{aligned}
m_{\widetilde{l}}=O(m_z)
\label{aba:eq3.2}
\end{aligned}
\end{split}
\end{equation}

\noindent where $\widetilde{q}$ can be any one of the scalar quarks (\ref{aba:eq2.3}) and $\widetilde{l}$ any of the scalar leptons (\ref{aba:eq2.4}).

These mass limits (\ref{aba:eq3.1}) and (\ref{aba:eq3.2}) are very strong constraints on the masses of the scalar quarks and scalar leptons.

\section{Masses of scalar quarks and scalar leptons}\label{sec4}

It only remains to put the information of the previous two sections together.

Consider first the twelve scalar quarks listed in (\ref{aba:eq2.3}). Since the scalar quarks have the same quantum number as the quarks except the spin, these scalar quarks have a charge of either $\frac{2}{3}$ or $-\frac{1}{3}$ just as the quarks, and hence they all couple to the $Z$. By the arguments of Sec. 2 and Sec. 3, their masses may be expected to be between $\frac{1}{2}m_z$ and very roughly $m_t \sim{2m_z}$:

\begin{equation}
\frac{1}{2} m_z < m_{\widetilde{q}} \lesssim 2 m_z
\label{aba:eq4.1}
\end{equation}

The corresponding argument for the scalar leptons is somewhat more complicated. An inequality similar to (\ref{aba:eq4.1}) holds for

\begin{equation}
\begin{split}
\begin{aligned}
\widetilde{\nu}_{eL}, \widetilde{e}_L, \widetilde{e}_R; \widetilde{\nu}_{\mu L}, \widetilde{\mu}_L, \widetilde{\mu}_R; \widetilde{\nu}_{\tau L}, \widetilde{\tau}_L, \widetilde{\tau}_R.
\label{aba:eq4.2}
\end{aligned}
\end{split}
\end{equation}

How about the remaining three: $\widetilde{\nu}_{eR}$, $\widetilde{\nu}_{\mu R}$, and $\widetilde{\nu}_{\tau R}$?

Since the masses of the scalar quarks and the scalar leptons are not very different from those of the known quarks and leptons in the sense of the inequality (\ref{aba:eq4.1}), it seems reasonable to entertain the possibility that the similarity between these particles goes even further. Specifically, since the known neutrinos have a lower mass (in fact a much lower mass) than the other quarks and leptons, it is hereby assumed that the three right-handed scalar neutrinos have significantly smaller masses than the other scalar quarks and scalar leptons.

It is not clear what is meant by the masses being "smaller", but they should be much smaller than the mass of the $Z$.

The picture that emerges from these considerations is the following:

1) These are twelve scalar quarks and twelve scalar leptons;

2) Of these twenty four scalar particles, twenty one have masses of the order of the mass of the $Z$; and

3) The other three scalar particles, the three right-handed scalar neutrinos, have small masses.

We are happy with the small messes of the neutrinos and the right-handed scalar neutrinos. In some sense, this seems to put even more content into the concept of the fermion-boson symmetry.

\section{Speculation on the decays of lightest scalar quark}\label{sec5}

Without introducing additional elementary particles, the lightest scalar quark must be stable, since it has nothing to decay into.

This is not consistent with the experimental data from the ATLAS Collaboration and the CMS collaboration at LHC~\cite{cmsatlas}. Suppose the lightest scalar quark is $\widetilde{q}$, where this $\widetilde{q}$ can be either charge $\frac{2}{3}$ or $-\frac{1}{3}$. Then there are, among others, the fermionic mesons, $\widetilde{q}$ $\overline{u}$ and $\widetilde{q}$ $\overline{d}$. Of these two fermionic mesons, one is charged while the other is neutral. Since $\widetilde{q}$ is stable, one of these two fermionic mesons is stable and the other is long lived~\cite{richard}. Therefore, the charged one of these two fermionic mesons is either stable or long lived, a result that contradicts the ATLAS and CMS data~\cite{cmsatlas}.

The lightest scalar quark must decay, and it is mandatory to introduce a new particle for it to decay into. Since there is no available experimental information on this decay, let us speculate on one of the simplest possibilities.

Since the lightest scalar quark has color, one of the decay products must carry this color; this means that one of the decay products is likely to be an ordinary quark. Since the scalar quark is a boson while the ordinary quark is a fermion, the other decay product, taken together, must be a fermion. Since these other decay products have neither color nor lepton number, a new elementary particle must be introduced.

Thus the simplest speculation is that a scalar quark can decay into an ordinary quark and a new fermion:
\begin{equation}
\begin{split}
\begin{aligned}
scalar\ quark \rightarrow quark + new\ fermion\ f,
\label{aba:eq5.1}
\end{aligned}
\end{split}
\end{equation}

\noindent where the new fermion may be real or virtual. This decay is available to all scalar quarks, including the lightest one.

This decay is in addition to the one analogous to that of quarks:

\begin{equation}
\begin{split}
\begin{aligned}
heavier\ scalar\ quark \rightarrow lighter\ scalar\ quark + W,
\label{aba:eq5.2}
\end{aligned}
\end{split}
\end{equation}

\noindent where again the $W$ may be real or virtual. Unlike (\ref{aba:eq5.1}), however, this decay (\ref{aba:eq5.2}) is not available to the lightest scalar quark.

Referring once more to the LEP measurement of the $Z$ width, these new fermions must again have masses larger than $\frac{1}{2}m_z$. Since these masses of the new fermions come from the Englert-Brout-Higgs mechanism~\cite{FERB}, some of the new fermion \textemdash the right-handed ones \textemdash are $SU(2)$ singlets, while others \textemdash the left-handed ones \textemdash are $SU(2)$ doublets.

Invoking once more the experimenal data~\cite{cmsatlas} from the ATLAS collaboration and the CMS Collaboration, we reach the conclusion that the charged new fermions again cannot be stable.

\section{Decays of the new fermions}\label{sec6}

How do the new fermions decay? There is even less information on these decays.

Clearly the decay products of the new fermion must contain at least one fermion. If this fermion in the decay products is one that is already introduced besides the new fermions themselves (namely the scalar quarks, the scalar leptons, and the known elementary particles in the standard model), then it must be either a quark or a lepton. If it is a quark, then the necessary coupling is essentially that of (\ref{aba:eq5.1}). Since this (\ref{aba:eq5.1}) has already been used for the decay of the scalar quark, it cannot be responsible for the decay of the new fermion.

Therefore in this case the fermion in the decay products must be a lepton. A moment's reflection then shows that the decay is 

\begin{equation}
\begin{split}
\begin{aligned}
new\ fermion\ f \rightarrow scalar\ lepton + \textit{anti-lepton}.
\label{aba:eq6.1}
\end{aligned}
\end{split}
\end{equation}

Since the mass of the new fermion $f$ is at least $\frac{1}{2}m_z$, this decay can always proceed if the scalar lepton is chosen to be one of the right-handed scalar neutrinos, i.e,

\begin{equation}
\begin{split}
\begin{aligned}
\textit{new fermion f} \rightarrow \textit{anti-lepton} + \textit{right-handed scalar neutrino},
\label{aba:eq6.2}
\end{aligned}
\end{split}
\end{equation}

\noindent It is possible that a new fermion $f$ decays into another new fermion $f^{'}$, and then this $f^{'}$ decays through (\ref{aba:eq6.2}).

Because of this (\ref{aba:eq6.1}), the scalar leptons have decays similar to those of the scalar quarks, namely,
\begin{equation}
\begin{split}
\begin{aligned}
\textit{scalar lepton} \rightarrow lepton + \textit{new fermion f},
\label{aba:eq6.3}
\end{aligned}
\end{split}
\end{equation}

\noindent in addition to 

\begin{equation}
\begin{split}
\begin{aligned}
\textit{heavier scalar lepton} \rightarrow \textit{lighter scalar lepton} + W,
\label{aba:eq6.4}
\end{aligned}
\end{split}
\end{equation}

Note that, in writing down (\ref{aba:eq6.1}) and (\ref{aba:eq6.2}), we have used the equality

\begin{equation}
\begin{split}
\begin{aligned}
\textit{the new fermions on the right-hand side of } (\ref{aba:eq5.1})\\
= \textit{the new fermions on the right-hand side of } (\ref{aba:eq6.3}).
\label{aba:eq6.5}
\end{aligned}
\end{split}
\end{equation}

There may or may not be additional elementary particles beyond those already introduced in Sec. 2 and 5. Even when there are, the decay (\ref{aba:eq6.2}) can still proceed.

What can these additional elementary particles be? These does not seem to be any experimental indication of what they may be. However, the following comment is perhaps appropriate.

The starting point of the present consideration is to use the concept of fermion-boson symmetry to introduce the scalar quarks and the scalar leptons into the standard model. There are many open issues, including the following two.

1) Should the fermion-boson symmetry used here be applied to other particles?

2) In particular, does the Higgs particle have any fermionic partner?

Because of our emphasis on the results from high-energy experiments, there is no way to answer these questions now within the present framework. Rather, these questions should be studied only after at least some of the scalar quarks and/or scalar leptons have been observed experimentally. Attention is therefore turned to experimental considerations.

\section{Experimental considerations}\label{sec7}

In order to observe the scalar quarks and the scalar leptons, the first step is to discover the new fermions $f$. Experimentally, it seems easier to observe the charged new fermions, say $f^+$, than the neutral ones. Written out more explicitly, the decays of $f^+$, according to (\ref{aba:eq6.2}), are

\begin{equation}
\begin{split}
\begin{aligned}
f^+ \rightarrow e^+ + \textit{right-handed scalar neutrino}
\label{aba:eq7.1}
\end{aligned}
\end{split}
\end{equation}

\begin{equation}
\begin{split}
\begin{aligned}
f^+ \rightarrow \mu^+ + \textit{right-handed scalar neutrino},
\label{aba:eq7.2}
\end{aligned}
\end{split}
\end{equation}

\noindent and

\begin{equation}
\begin{split}
\begin{aligned}
f^+ \rightarrow \tau^+ + \textit{right-handed scalar neutrino}.
\label{aba:eq7.3}
\end{aligned}
\end{split}
\end{equation}

\noindent In each of these cases, the right-handed scalar neutrino escapes the detector.

These decays bear some similarity to the leptonic decays of $W$, namely
\begin{equation}
\begin{split}
\begin{aligned}
W^+ \rightarrow e^+ + \nu_e,
\label{aba:eq7.4}
\end{aligned}
\end{split}
\end{equation}

\begin{equation}
\begin{split}
\begin{aligned}
W^+ \rightarrow \mu^+ + \nu_\mu,
\label{aba:eq7.5}
\end{aligned}
\end{split}
\end{equation}

\noindent and

\begin{equation}
\begin{split}
\begin{aligned}
W^+ \rightarrow \tau^+ + \nu_\tau,
\label{aba:eq7.6}
\end{aligned}
\end{split}
\end{equation}

In each of these decays (\ref{aba:eq7.1})-(\ref{aba:eq7.6}), one of the decay products cannot be detected, either a known neutrino or an unknown right-handed scalar neutrino. The decays (\ref{aba:eq7.4}) and (\ref{aba:eq7.5}) have been of central importance both in the original discovery of the $W$ by the $UA1$ Collaboration and the $UA2$ Collaboration~\cite{UA1} and later in the search for a second $W$, usually called $W^{'}$. 

This similarity is, however, overshadowed by the following fundamental difference between the production of $W^{'}$ and that of the charged new fermion at the Large Hadron Collider. While the $W^{'}$ is typically produced singly, the new fermions are predominately pair produced. A significant fraction consists of a pair $f^{+}$ and $f^{-}$; their leptonic decays are characterized by the pair $e^{+}e^{-}$, $\mu^+\mu^-$, $e^{+}\mu^{-}$, or $\mu^{+}e^{-}$. Such lepton pairs, both with high transverse momentum, can be used as a basis of searching for the scalar quarks and scalar leptons at the Large Hadron Collider.

The proposed sequence of experiments is therefore as follows.

Step \Romannum{1}: The search for the charged new fermion. This search is to be carried out by analyzing events where there is a pair of opposite charged high-$p_T$ leptons $e^+e^-$, $\mu^+\mu^-$, $e^+\mu^-$, or $\mu^+e^-$ coming from $f^+f^-$.

Step \Romannum{2}: The search for scalar quarks and scalar leptons. After the new fermions have been observed experimentally as described in Step \Romannum{1}, the search for scalar quarks and scalar leptons can begin. These searches will be based on the decays (\ref{aba:eq5.1}) and (\ref{aba:eq6.3}) where in each case one of the decay products is the new fermion.

It should be noted that, when the equality (\ref{aba:eq6.5}) holds so that the new fermions in the decays (\ref{aba:eq5.1}) and (\ref{aba:eq6.3}) are the same, the above Step \Romannum{2} can lead to both the scalar quarks and the scalar leptons. If the new fermions in the decays (\ref{aba:eq5.1}) and (\ref{aba:eq6.3}) are not the same, the above Step \Romannum{2} can still lead to the scalar leptons.

\section{Back to quantum field theory}\label{sec8}

The basic theme of the present paper is to combine the concept of fermion-boson symmetry with the desire to reduce divergences in quantum field theory. This combination leads to consequences that do not follow from either one alone. The results as applied to the properties of scalar quarks and scalar leptons have been studied in Secs. 4-7; it remains to discuss the possible implications on gauge quantum field theory, i.e., Yang-Mills non-Abelian gauge theory~\cite{CNYRL}.

In general, a renormalizable quantum field theory has both quadratic divergences and logarithmic divergences in the perturbation series. [As an exception, there is no quadratic divergence in quantum electrodynamics.] Renormalization is one of the greatest theoretical achievements of the twentieth century~\cite{schwinger}. It is a most beautiful theory that makes quantum electrodynamics the first successful field theory with amazing predictive power.

Nevertheless, it must be admitted that a quantum field theory without any divergence in its perturbation series is even better. This is the motivation for the papers in Ref ~\cite{ferrara}, where the cancellation of quadratic divergences is found to give information on the particle masses.

In this context, we would like to make the following conjecture. With the fermion-boson symmetry introduced into the standard model in a suitable way, all divergences, both quadratic and logarithmic, can be cancelled. In this way, it will be possible to get a finite Yang-Mills non-Abelian gauge theory. The requirement of having such a finite theory will dictate how the additional particles can be introduced.

It is hoped that such a finite quantum field theory can be developed in the not-too-distant future.

\section{Discussion}\label{sec9}

Whenever the fermion-boson symmetry is mentioned, supersymmetry comes to mind. The theory of supersymmetry, which is based on the idea of graded Lie algebra and is by now well developed, is a beautiful mathematical theory. As physics, however, it suffers from a lack of contact with results of high-energy experiments, and therefore a lack of theoretical predictions.

In views of this situation, it is the purpose here to investigate this concept of fermion-boson symmetry with emphasis on data from high-energy experiments. This implies immediately that the starting point must be the standard model with the addition of scalar quarks and scalar leptons. In other words, this fermion-boson symmetry is applied initially in a more limited sense only to quarks and leptons, but not to the gauge and the Higgs particles. The experimental data from the electron-positron collider LEP are used to give some constraints on the properties of these scalar quarks and scalar leptons.

Further information on these scalar particles are obtained from two different considerations. On the theoretical side, the desire to reduce divergences in quantum field theory gives further constraints on the masses of these scalar particles. On the experimental side, the data from the Large Hadron Collider indicate the lightest scalar quark should not be stable, leading to new fermions for it to decay into. A method is proposed to observe these new fermions at the Large Hadron Collider. After these new fermions are discovered, then it seems feasible to find the scalar quarks and the scalar leptons.

 Here is a succinct comparison of supersymmetry with the present approach.
 
1) While supersymmetry is a well defined theory, the present approach is still being developed: in particular, even the particle content is incomplete at present, waiting for further experimental data from the Large Hadron Collider.

2) Graded Lie algebra plays a central role in the theory of supersymmetry, but is not involved in the present approach.

3) In the present approach, there are, for each generation, four different masses for the scalar quarks and four different masses for the scalar leptons -- see Sec. 2.  Therefore, soft supersymmetry breaking is not needed.  In other words, in the Lagrangian density (which is still incomplete), every term is of dimension four except the one mass term (that of the Higgs doublet), just like the standard model~\cite{sglashow} and quantum electrodynamics.

4) In this way, the introduction of the 105 independent coupling constants for the soft supersymmetry breaking terms in the theory of supersymmetry is avoided.

5) On the phenomenological side, a distinctive feature for the present approach is that the masses of the right-handed scalar neutrinos, one of which may be identified as the lightest supersymmetric particle, are much less than the mass of the $Z$ -- see Sec. 4.  Such a small mass is not inconsistent with supersymmetry.

6) As a consequence of this small mass, the method of searching for the scalar quarks and the scalar leptons is different from the previous searches -- see Sec. 7.

\section{Acknowledgements}

We are grateful to the hospitality at CERN, where part of this work has been carried out.


\end{document}